\documentclass{PoS}
\usepackage{alltt,feynarts,graphicx,amsmath,amssymb,colordvi,axodraw}

\newcommand{\eg}{e.g.\ }
\newcommand{\ie}{i.e.\ }

\newcommand\lbrac{\symbol{123}}
\newcommand\rbrac{\symbol{125}}
\newcommand\Brac[1]{\lbrac#1\rbrac}

\newcommand{\ket}[1]{\left| #1\right\rangle}
\newcommand{\bra}[1]{\left\langle #1\right|}
\newcommand{\braket}[2]{\left\langle #1\vphantom{#2}%
  \right. \kern-2.5pt\left| #2\vphantom{#1}\right\rangle}
\newcommand\icon[1]{\lower 4pt\hbox{\includegraphics[scale=.6]{#124}}}
\newcommand\Code[1]{\ensuremath{\texttt{#1}}}
\newcommand\Var[1]{\ensuremath{\mathit{#1}}}

\title{Extensions in FormCalc 5.3}

\ShortTitle{Extensions in FormCalc 5.3}

\author{\speaker{T.~Hahn} \\
	Max-Planck-Institut f\"ur Physik, \\
	F\"ohringer Ring 6, D--80805 Munich, Germany}

\author{J.I.~Illana \\
	Departamento de F{\'\i}sica Te\'orica y del Cosmos, and
	Centro Andaluz de F{\'\i}sica de Part{\'\i}culas
	Elementales (CAFPE), \\
	Universidad de Granada, \\
	E--18071 Granada, Spain}

\abstract{We present a new tool for editing Feynman diagrams as well as
several extensions in version 5.3 of the package FormCalc for the 
calculation of Feynman diagrams.}

\FullConference{XI International Workshop on Advanced Computing and Analysis Techniques in Physics Research \\
		 April 23--27 2007 \\
		 Amsterdam, the Netherlands}

\begin{document}


\section{Introduction}

FeynArts \cite{FeynArts} and FormCalc \cite{FormCalc} are Mathematica
packages for the generation and calculation of Feynman diagrams up to
one-loop order.  This contribution describes a new tool for graphically
editing Feynman diagrams in FeynArts' \LaTeX\ format as well as several
new features of FormCalc:
\begin{itemize}
\item A Mathematica interface for FormCalc-generated Fortran code.
\item The splitting of abbreviations into tree and loop parts (this can 
      greatly affect performance).
\item The implementation of four-dimensional Fierz identities.
\item A new function to express amplitudes in terms of phase-space 
      variables fully analytically.
\item New functions for easier definition of renormalization constants.
\item A separate diagonalization package.
\end{itemize}


\section{Editing Feynman Diagrams}

FeynEdit is a Java program for editing Feynman diagrams.  It uses the
\LaTeX\ representation of FeynArts \cite{FeynArts} for input and output. 
Diagrams are entered into and retrieved from the editor through
cut-and-paste with the mouse.  This makes it unnecessary to first save
the diagrams one wants to edit in a separate file.

Changing the geometry of a diagram has thus become easy.  At least in
the present version, however, the editor does not show details such as
line styles and the actual labels.  This is mostly for performance
reasons (think of dragging a gluon line).  With the FeynArts \LaTeX\
format, it is not difficult to change these elements later in a text 
editor.

\begin{figure}  
\centerline{\includegraphics{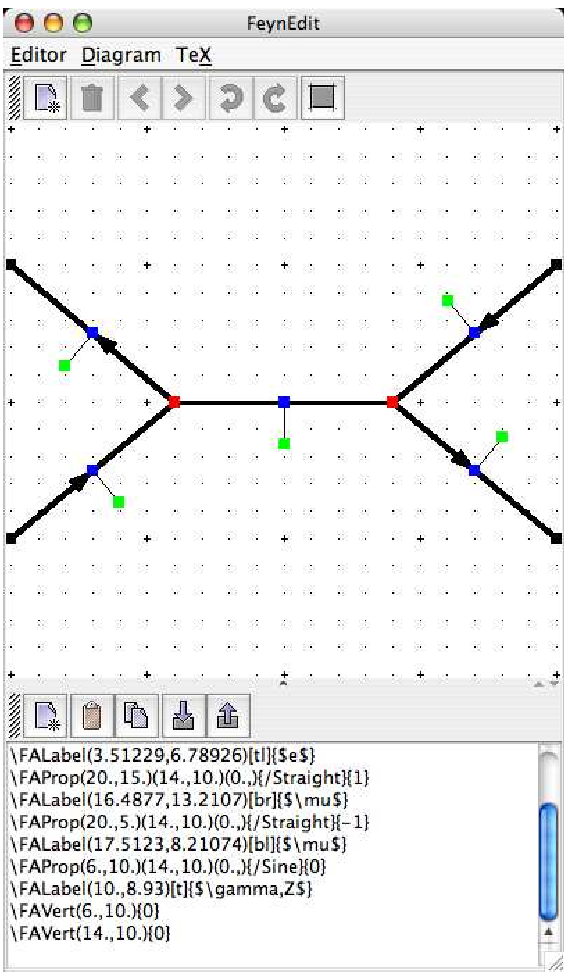}\qquad\includegraphics{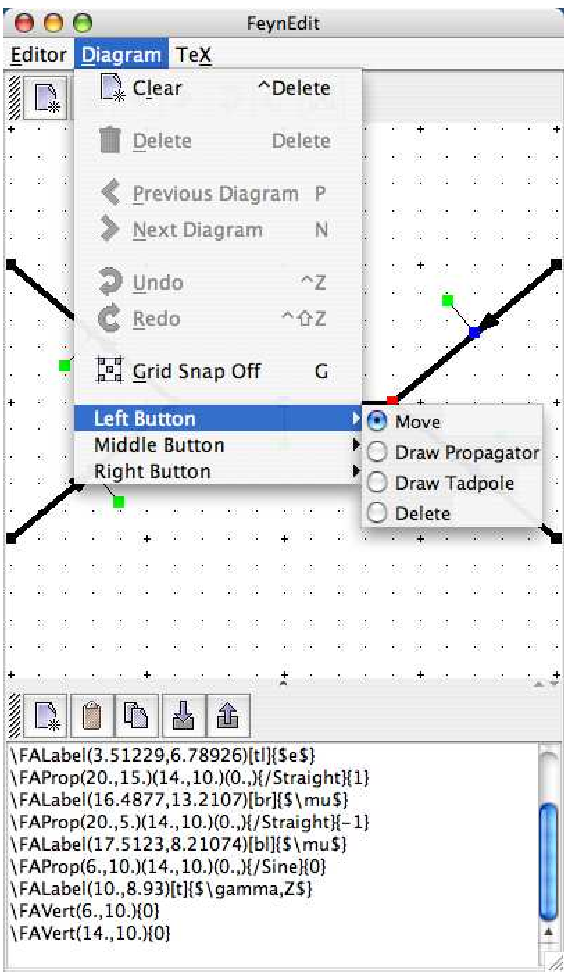}}
\caption{\label{fig:diagram}Left panel: The diagram pasted into FeynEdit
and displayed.  Right panel: The Mouse Button Assignment Menu in 
FeynEdit.}
\end{figure}

The window is divided into an upper panel for the diagram display and a
lower panel which shows the \LaTeX\ code (see Fig.~\ref{fig:diagram}). 
To display an existing Feynman diagram, mark its \LaTeX\ code with the
mouse and paste it into the lower dialog box.  Then press the
\icon{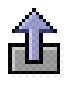}\,button to display the diagram.  Otherwise, start with an
empty canvas and use the mouse to add elements.

When finished with editing, press the \icon{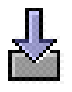}\,button to turn the
diagrams into \LaTeX\ code, then pick up the latter with the mouse and
paste it (back) into your text.

The diagram can be edited with the mouse.  Four editing functions are
available:
\begin{itemize}
\item Move vertices, propagators, and labels:
      Click on the corresponding box (red, blue, green) and drag
      it to the desired position.

\item Draw tadpoles:
      Click on the `footpoint' of the tadpole and drag it to the
      desired size and orientation.

\item Draw `ordinary' propagators:
      Click on the starting point and drag to the end point.

\item Delete objects:
      Click on the square (red, blue, green) corresponding to the
      object you want to delete.  When deleting a vertex, the
      propagators adjacent to this vertex are also deleted.  When
      deleting a propagator, the corresponding label is also deleted.
\end{itemize}
In the default setup, the left mouse button moves objects, the middle
mouse button draws tadpoles, and the right mouse button draws
propagators.  The assignment of the mouse button can be changed in the
Mouse Button Assignment menu (Fig.~\ref{fig:diagram}, right panel).


\section{Mathematica Interface}

The new Mathematica Interface turns the generated stand-alone Fortran
code into a Mathematica function for evaluating the cross-section or
decay rate as a function of user-selected model parameters.  The
benefits of such a function are obvious, as the whole instrumentarium of
Mathematica commands can be applied to them.  For example, one can
simply use the Mathematica function \Code{ContourPlot} to produce a
contour plot of the cross-section.

Interfacing is done using the MathLink protocol.  It is important to
realize that the cross-section is not evaluated in Mathematica, but in
Fortran, and only the numerical results computed by the Fortran code are
transferred back to Mathematica.


\subsection{Input}

The changes necessary to produce a MathLink executable are by design
minor and affect only the file \Code{run.F}, where the user has to
choose which model parameters are interfaced from Mathematica.  A
typical line in the stand-alone version of \Code{run.F} might look like:
\begin{verbatim}
   #define LOOP1 do 1 TB = 5, 50, 5
\end{verbatim}
This declares a loop over the variable \Code{TB} (= $\tan\beta$ in the 
MSSM model file) running from 5 to 50 in steps of 5.
To turn this code into a Mathematica program, with \Code{TB} appearing
as an argument of the Mathematica function, the only modification
necessary is to change the above line into
\begin{verbatim}
   #define LOOP1 call MmaGetReal(TB)
\end{verbatim}
The variable \Code{TB} is `imported' from Mathematica now, \ie the
cross-section function in Mathematica becomes a function of \Code{TB}
hereby.

The user has full control over which variables are `imported' from
Mathematica and which are set in Fortran.  Specifically, the invocations
of \Code{MmaGetReal} and its companion subroutine \Code{MmaGetComplex}
serve two purposes.  At compile time they determine with which arguments
the Mathematica function is generated, and at run time they actually
transfer the function's arguments to the specified Fortran variables.

Once the makefile detects the presence of these subroutines, it
automatically generates interfacing code and compiles a MathLink
executable.  For a file \Code{run.F} the corresponding MathLink
executable is also called \Code{run}, as in the stand-alone case.


\subsection{Output}

Similar to the \Code{MmaGetReal} invocations, the Fortran program can 
also `export' variables to Mathematica.  The line that prints a 
parameter in the stand-alone code is of the form
\begin{verbatim}
   #define PRINT1 SHOW "TB", TB
\end{verbatim}
To transmit the value of \Code{TB} to Mathematica, this becomes
\begin{verbatim}
   #define PRINT1 call MmaPutReal("TB", TB)
\end{verbatim}


\subsection{Usage}

Once the changes to \Code{run.F} are made, the program \Code{run} is 
compiled as usual:
\begin{verbatim}
   ./configure
   make
\end{verbatim}
It is then loaded in Mathematica with
\begin{verbatim}
   Install["run"]
\end{verbatim}
which makes a Mathematica function of the same name, \Code{run},
available.  There are two ways of invoking it which correspond closely 
to the command-line invocation of the stand-alone executable:
\begin{itemize}
\item Compute a differential cross-section at $\sqrt s = \Code{sqrtS}$:
\begin{verbatim}
   run[sqrtS, arg1, arg2, ...]
\end{verbatim}
\item Compute a total cross-section for $\Code{sqrtSfrom}\leqslant
\sqrt s\leqslant\Code{sqrtSto}$:
\begin{verbatim}
   run[{sqrtSfrom, sqrtSto}, arg1, arg2, ...]
\end{verbatim}
\end{itemize}
The extra arguments \Code{arg1}, \Code{arg2}, \dots\ are precisely the 
variables imported from the Fortran code, such as \Code{TB} in the
example above.


\subsection{Data Retrieval}

Both the parameters exported from the Fortran code and the computed data
(cross-section, decay rate, etc.) are transferred in sets, meaning that
a separate Mathematica definition is made for each set.  This has the
important advantage that if the calculation is prematurely aborted,
parameters and data transferred so far are still accessible.  Such sets
might look like
\begin{small}
\begin{verbatim}
   Para[1] = {TB -> 5., MA0 -> 250.}
   Data[1] = {DataRow[{500.}, {0.0539684, 0.}, {2.30801 10^-21, 0.}],
              DataRow[{510.}, {0.0515943, 0.}, {4.50803 10^-22, 0.}]}
\end{verbatim}
\end{small}
Not surprisingly, the actual return value of the function \Code{run} 
is (only) an integer which indicates how many sets have been 
transferred.

The \Code{Para} sets contain the parameters exported from the Fortran 
code.  The \Code{Data} sets contain for every data point computed a 
\Code{DataRow} object which has three arguments:
\begin{itemize}
\item the unintegrated kinematical variables, \\
here \eg \verb={500.}= $= \{\sqrt s\}$,

\item the tree-level cross-section and one-loop correction, \\
here \eg \verb={0.0539684, 0.}= $=
\{\sigma_{\text{tot}}^{\text{tree}}, 
\sigma_{\text{tot}}^{\text{1-loop}}\}$, and

\item the integration errors of these quantities, \\
here \eg \verb={2.30801 10^-21, 0.}= $=
\{\Delta\sigma_{\text{tot}}^{\text{tree}}, 
\Delta\sigma_{\text{tot}}^{\text{1-loop}}\}$.
\end{itemize}


\section{Splitting Abbreviations}

As a side effect of FormCalc's abbreviationing technique, the evaluation
of the abbreviations constitute a major part of an amplitude
calculation.  Timings for a one-loop calculation may very roughly look
like
\begin{center}
\begin{picture}(290,140)
\SetColor{White}
\SetWidth{2}
\ArrowLine(50,70)(50,55)
\ArrowLine(50,35)(50,20)
\SetWidth{1}
\Text(50,135)[]{\PineGreen{OLD}}
\CBox(0,70)(100,125){Blue}{PastelBlue}
\Text(50,97)[]{\Blue{Compute abbr}}
\CBox(0,35)(100,55){Red}{PastelRed}
\Text(50,45)[]{\Red{Compute $\mathcal{M}^{\text{tree}}$}}
\CBox(0,0)(100,20){Red}{PastelRed}
\Text(50,10)[]{\Red{Compute $\mathcal{M}^{\text{1-loop}}$}}
\SetOffset(110,0)
\SetWidth{2}
\ArrowLine(50,105)(50,90)
\ArrowLine(50,70)(50,55)
\ArrowLine(50,35)(50,20)
\SetWidth{1}
\Text(50,135)[]{\PineGreen{NEW}}
\CBox(0,105)(100,125){Blue}{PastelBlue}
\Text(50,115)[]{\Blue{Compute abbr$^{\text{tree}}$}}
\CBox(0,70)(100,90){Blue}{PastelBlue}
\Text(50,80)[]{\Blue{Compute abbr$^{\text{1-loop}}$}}
\CBox(0,35)(100,55){Red}{PastelRed}
\Text(50,45)[]{\Red{Compute $\mathcal{M}^{\text{tree}}$}}
\CBox(0,0)(100,20){Red}{PastelRed}
\Text(50,10)[]{\Red{Compute $\mathcal{M}^{\text{1-loop}}$}}
\SetOffset(170,0)
\Text(50,135)[l]{\PineGreen{CPU-time (rough)}}
\def\brc{$\left.\vrule width 0pt height 1ex depth 1ex\right\}$~}
\Text(50,115)[l]{\brc 5\,\%}
\Text(50,80)[l]{\brc 95\,\%}
\Text(50,45)[l]{\brc .1\,\%}
\Text(50,10)[l]{\brc .1\,\%}
\end{picture}
\end{center}
The old design had the obvious disadvantage that evaluating only the
tree-level part would take about as much CPU time as the full one-loop
amplitude.  The current version thus splits the abbreviations into those
that are needed for the tree-level part and the rest, and the main 
subroutine \Code{SquaredME} correspondingly has an additional flag to
choose whether to compute only the tree-level part.

The present set-up of the Fortran driver code does not (yet) make use of
this splitting automatically.  The most obvious application would be a
separate phase-space integration of tree-level and one-loop component.


\section{Fierz Identities in four dimensions}

The Fierz identities rearrange fermion chains by switching spinors, \ie
$$
\bra{1}\Gamma_i\ket{2} \bra{3}\Gamma_j\ket{4}
= \sum c_{kl} \bra{1}\Gamma_k\ket{4} \bra{3}\Gamma_l\ket{2}
$$
This is important in particular if one wants to extract certain
predefined structures from the amplitude, most notably Wilson
coefficients.

FormCalc so far automatically used the Fierz identities on 2-dimensional
(Weyl) spinors, where the application is straightforward.  The latest 
FormCalc version offers also the 4-dimensional variant through the 
new \Code{FermionOrder} option for the \Code{CalcFeynAmp} function.  It 
is used as in
\begin{verbatim}
   CalcFeynAmp[..., FermionChains -> Chiral,
                    FermionOrder -> {2, 1, 3, 4}]
\end{verbatim}
The \Code{FermionChains} option chooses 4-dimensional (Dirac) spinors
and the \Code{FermionOrder} option instructs \Code{CalcFeynAmp} to try
to bring the spinor chains into the order $\bra{2} X\ket{1} \bra{3}
Y\ket{4}$.


\section{Fully Analytic Amplitudes}

The `smallest' object appearing in the output of \Code{CalcFeynAmp} is a
four-vector.  The components, which reflect a particular phase-space
parameterization, are inserted only later, usually in the numerical
part.

So far, only the squared matrix element could be computed fully
analytically.  This method avoids an explicit phase-space
parameterization but has the disadvantage that the size of the squared
matrix element grows quadratically with the size of the amplitude.  For
example, for a fermionic amplitude $\mathcal{M} = \sum^N c_i F_i$, where
the $F_i$ are (products of) fermion chains, the squared matrix element
is computed as $|\mathcal{M}|^2 = \sum^{N^2} c_i c_j^* (F_i F_j^*)$.

The new add-on package VecSet (loaded with
\verb|<< FormCalc`tools`VecSet`|) makes two new functions available with 
which is is possible to express an amplitude in terms of phase-space 
parameters fully analytically:
\begin{itemize}
\item In a first step, the external vectors need to be set with the
\Code{VecSet} function.  For example, \Code{VecSet[1, m1, p1,
\Brac{0,0,1}]} sets the momentum, polarization vectors, and spinors for
external particle \#1 with mass \Code{m1} and three-momentum \Code{p1} 
moving in the direction \Code{\Brac{0,0,1}}.  This function is very 
similar to its Fortran namesake in the FormCalc drivers library.

\item Once all external vectors have been set, an amplitude \Code{amp} 
can be evaluated for example with \Code{ToComponents[amp, "+-+-"]}, 
where the \Code{"+-+-"} are the polarizations of the external particles.
\Code{ToComponents} delivers an expression in terms of the phase-space 
parameters used in \Code{VecSet}.
\end{itemize}


\section{New Functions for Renormalization Constants}

New functions which simplify the definition of
renormalization constants (RCs) have been introduced (the precise 
definitions of these quantities are listed in the FormCalc manual):
\begin{itemize}
\item \Code{MassRC[\Var{f}]}
	-- the mass RC $\delta M_{\Var{f}}$,
\item \Code{MassRC[\Var{f_1},\,\Var{f_2}]}
	-- the mass RC $\delta M_{\Var{f_1}\Var{f_2}}$,
\item \Code{FieldRC[\Var{f}]}
	-- the field RC $\delta Z_{\Var{f}}$,
\item \Code{FieldRC[\Var{f_1},\,\Var{f_2}]} --
	-- the field RC $\delta Z_{\Var{f_1}\Var{f_2}}$,
\item \Code{TadpoleRC[\Var{f}]}
	-- the tadpole RC $\delta T_{\Var{f}}$,
\item \Code{WidthRC[\Var{f}]}
	-- the width $\Gamma_f$.
\end{itemize}
Using these function, the entire renormalization section of the Standard
Model now fits on half a page (see FeynArts' \Code{SM.mod}).  The main
aspect is to avoid mistakes, however: for example, a subtle index error
in the fermion mixing counterterm in \Code{SM.mod} has been discovered
through the new RC functions.


\section{Separate Diagonalization Package}

\begin{figure}
\begin{center}
\includegraphics[width=.8\hsize]{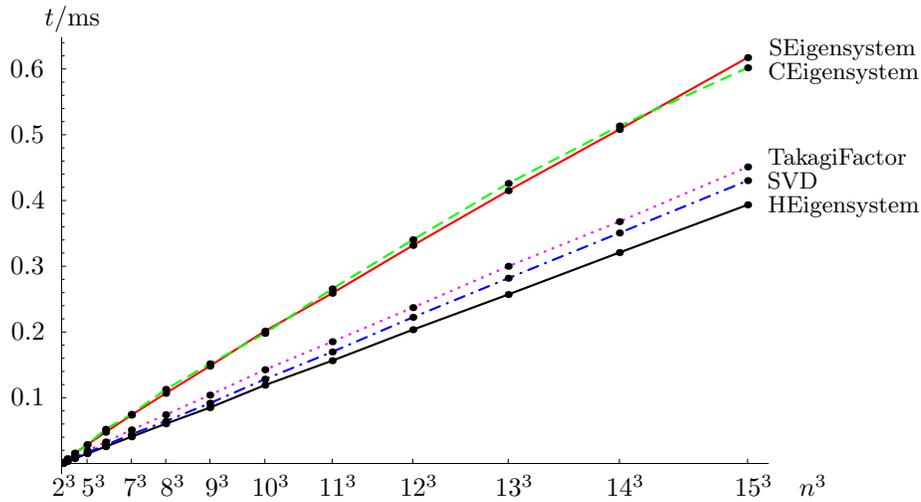}
\end{center}
\caption{\label{fig:timings}Timings of the diagonalization routines on
an AMD X2-5000.}
\end{figure}

The diagonalization routines included in FormCalc have been extended and 
made available as a separate package \cite{Diag}.  The following 
routines are available:
\begin{itemize}
\item \Code{HEigensystem} diagonalizes a Hermitian matrix,
\item \Code{SEigensystem} diagonalizes a complex symmetric matrix,
\item \Code{CEigensystem} diagonalizes a general complex matrix,
\item \Code{TakagiFactor} computes the Takagi factorization 
      of a symmetric matrix (\eg the neutralino mass matrix),
\item \Code{SVD} performs the Singular Value Decomposition.
\end{itemize}
All routines are based on the Jacobi algorithm.  This is conceptually
simple but scales less favourably than \eg the QR method.  The
applicability range is thus small to medium-size matrices.  See also the 
timings in Fig.~\ref{fig:timings}.

Use of the Jacobi algorithm results in rather compact code ($\sim$ 3
kBytes each), which is therefore easy to adapt to own conventions.  The
library is implemented in Fortran 77, but has a C/C++ and Mathematica
interface.  It stands under the LGPL license.

\section{Summary and Availability}

\begin{itemize}
\item
The drawing tool FeynEdit is available from http://www.feynarts.de
as an extra package (not part of FeynArts).

\item
The FormCalc version 5.3 contains the main new features
\begin{itemize}
\item Mathematica interface,
\item Abbreviations are split into tree-level and loop parts,
\item Fierz identitites in four dimensions are implemented,
\item Fully analytic amplitudes are possible through an add-on package,
\item New functions for renormalization constants
\end{itemize}
and is available from http://www.feynarts.de/formcalc.

\item
FormCalc's diagonalization routines have been extended and modeled into 
an own package which is available from http://www.feynarts.de/diag.
\end{itemize}

\end{document}